\documentclass[journal]{IEEEtran}
\usepackage{cite}

\ifCLASSINFOpdf
\else
   \usepackage[dvips]{graphicx}
\fi
\usepackage[cmex10]{amsmath}
\usepackage{amsfonts}
\usepackage[tight,footnotesize]{subfigure}
\begin{document}
%
\title{On the Distributed Construction of a Collision-Free Schedule in WLANs}
%
%
%

\author{
	Jaume~Barcelo,~
    Azadeh~Faridi,~
    Boris~Bellalta,~
    Gabriel~Martorell,~
    and~David~Malone
\thanks{J. Barcelo, Azadeh Faridi and Boris Bellalta are with Universitat Pompeu Fabra.\newline
G. Martorell is with Universitat de les Illes Balears.\newline
D. Malone is with the Hamilton Institute, National University of Ireland.\newline
The corresponding author is J. Barcelo. E-mail address: jaume.barcelo@upf.edu Postal Address: Roc Boronat, 138, 08018, Barcelona.}%
}

\maketitle

\begin{abstract}
In wireless local area networks (WLANs), a media access protocol arbitrates access to the channel.
In current IEEE 802.11 WLANs, carrier sense multiple access with collision avoidance (CSMA/CA) is used.
Carrier sense multiple access with enhanced collision avoidance (CSMA/ECA) is a subtle variant of the well-known CSMA/CA algorithm that offers substantial performance benefits.
CSMA/ECA significantly reduces the collision probability and, under certain conditions, leads to a completely collision-free schedule.
The only difference between CSMA/CA and CSMA/ECA is that the latter uses a deterministic backoff after successful transmissions.
This deterministic backoff is a constant and is the same for all the stations.

The first part of the paper is of tutorial nature, offering an introduction to the basic operation of CSMA/ECA and describing the benefits of this approach in a qualitative manner.
The second part of the paper surveys related contributions, briefly summarizing the main challenges and potential solutions, and also introducing variants and derivatives of CSMA/ECA.

\end{abstract}


%
\IEEEpeerreviewmaketitle

\section{Introduction}
%
%
%
%
\IEEEPARstart{T}{he} distributed sharing of a medium by multiple stations is a classic communications problem. 
The AlohaNet network \cite{abramson2009asw} pioneered the use of random protocols as media access control (MAC) protocols.
This network connected several wireless stations in different islands of the Hawaiian archipelago.
The MAC protocol that was used there is known as the Aloha protocol and is very simple.
A wireless station transmits when it has a packet to be transmitted. 
If the transmission fails, the transmission is reattempted after a random backoff time.

A particularity of random access protocols is the possibility of collisions.
A collision occurs when multiple stations access the medium simultaneously and their transmissions cannot be correctly decoded.
These collisions can be resolved by means of retransmissions, but they increase the delay and reduce the maximum throughput of the network.

Despite collisions being detrimental for the network performance, Aloha is still an interesting option for channel access because it exhibits some key properties.
The first one is its distributed nature, since Aloha does not require any central entity to operate.
Aloha is also very simple and easy to implement.
Furthermore, it is also extremely robust as it can quickly recover from network problems such as a short interference burst.
Finally, the Aloha protocol does not require a heavy signaling overhead.
The combination of these properties was the fundamental reason for the success of Aloha and similar protocols that followed.

The original (or classic) Aloha protocol suffered from some inefficiencies when it had to deal with moderate to high traffic loads.
For this reason, the original protocol was followed by other derivatives that introduced some refinements or adjustments for a particular network or traffic pattern.
Two of these derivatives are Slotted Aloha and Reservation Aloha.

The Slotted Aloha protocol divides the time into fixed length slots and the stations can transmit only at the beginning of those slots.
In doing so, it decreases the chances of collisions and, under the assumption of fixed packet length that perfectly fits into the aforementioned fixed slot length, it doubles the throughput of the original Aloha.

Reservation Aloha \cite{crowther1973sbc,tasaka1983spr} extends Slotted Aloha with a reservation mechanism.
Reservation Aloha was intended for random access satellite communications and assumes that the stations listen continuously to the channel and can distinguish between occupied and free slots.
In  Reservation Aloha, a number of consecutive  time slots are grouped in a frame.
The frames are useful in the reservation process and they all contain the same number of slots.
When a station successfully transmits in a given slot of a frame, it implicitly makes a reservation on the same slot of the following frame.
This reservation can be very advantageous in some networks, increasing the capacity and reducing the delay.
In Reservation Aloha, different nodes implicitly agree on a collision-free schedule that results in a better utilization of the shared channel.
However, Reservation Aloha also introduces some complexities, such as choosing the right frame size or handling the situations in which there are more stations than available slots in a frame.
Both Slotted Aloha and Reservation Aloha use fixed-sized slots.

The original Aloha network was designed for inter-island communication.
The Slotted Aloha and the Reservation Aloha protocols were used in satellite communications.
Wireless Local Area Networks (WLANs) represent a different scenario because the distances are much shorter.
When the propagation times are short compared to the duration of the transmission of a packet, and all the stations can hear each other's transmissions, empty slots can be made much shorter than busy slots.
This way the network performance can be improved, since the channel will be idle for a smaller fraction of time and therefore, there will be more time for successful transmissions.
It is possible to shorten the empty slots in this case because, when stations are close to each other, they can quickly detect whether a slot is busy or empty by simply sensing the medium at the beginning of each slot.
This technique is called Carrier Sense Multiple Access (CSMA) and is used in WLANs that implement the IEEE 802.11 standard.
IEEE 802.11 WLANs are popularly known by their certification name, WiFi, and are prevalent in the marketplace.
It is not the goal of this paper to delve into the details of the standard, but we will use it as a reference CSMA implementation.

Recently, different research initiatives have suggested to combine the advantages of Reservation Aloha and CSMA (e.g., \cite{barcelo2008lba,he2009srb,barcelo2010fcc,fang2011dlm,barcelo2011tcf,martorell2012pec}). 
The goal is to construct, in a distributed way, a collision-free schedule that repeats periodically.
This schedule consists of some short empty slots and some long successful slots.
The novelty of our proposed protocol is that long collision slots are avoided, thus substantially increasing the network throughput.
The fact that the participating stations transmit in a round-robin fashion offers good jitter and fairness properties.

The motivation of this paper is to offer an introduction to this new family of protocols and provide an overview of recent work in this area.
We will describe CSMA/ECA (where ECA stands for Enhanced Collision Avoidance) as an example of a contention protocol that uses a deterministic backoff after successful transmissions to reduce the number of collisions.
Then we will review different related contributions to discuss  the research challenges, the performance gains, and the scenarios of interest and applicability.
We will also briefly mention the generality of the underlying principle and the possibility of using it for diverse problems in the field of radio resource management.

The remainder of the paper is organized as follows.
The next section provides a tutorial on CSMA/ECA.
Then, in Section~\ref{sec:survey} we summarize a selection of papers in this research area to provide an overview of the state of the art. 
The paper ends with some concluding remarks in Section~\ref{sec:conclusion}.

\section{Collision avoidance (CA) and enhanced collision avoidance (ECA) in CSMA networks}
\label{sec:eca}

\begin{figure*}[!t]
\centering
\subfigure[CSMA/CA]{\includegraphics[width=5.5in]{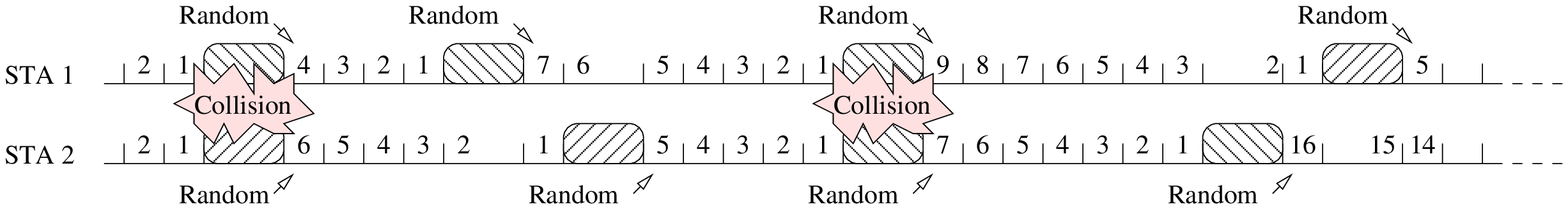}%
\label{fig:csma_ca}}
\subfigure[CSMA/ECA]{\includegraphics[width=5.5in]{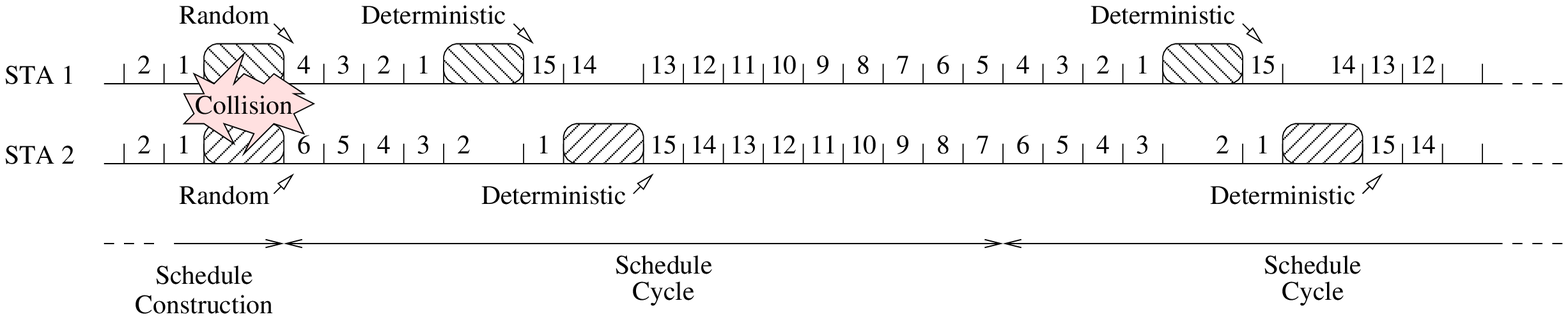}%
\label{fig:csma_eca}}
\caption{Examples of contention in which two wireless stations compete for channel access. The rounded boxes represent transmissions and the numbers are the backoff counters. It can be observed that CSMA/ECA attains cyclic collision-free operation after the construction of the schedule (transient convergence).}
\label{fig:ca_vs_eca}
\end{figure*}

Most of the currently deployed WLANs are compliant with the IEEE 802.11 standard and rely on CSMA/CA to share the channel time.
Thanks to the carrier sense capabilities of the CSMA stations, channel time can be divided in variable length slots.
We classify the slots as either empty, if no station transmits, or busy, if one or more stations transmit.
Among busy slots, we differentiate between successful slots, when there is a single transmission, and collision slots, when multiple stations simultaneously transmit.
Empty slots are relatively short and of constant duration, which is specified by the standard, and  busy slots are of variable length. 
As an example, the empty slot duration for IEEE 802.11b is 20 $\mu$s and a busy slot can be 1200 $\mu$s long.

Since the stations can use carrier sensing to detect the end of a transmission, it is possible to synchronize the nodes to the end of variable length transmissions.
The fact that the empty slots can be orders of magnitude shorter than the busy slots represents a performance gain over those approaches in which the slot size is fixed and constant.

In wired networks, it is possible for the nodes involved in a collision to detect the collision while it is taking place and immediately stop transmitting.
This technique is called CSMA with collision detection (CSMA/CD) and keeps the duration of collision slots very short.

In contrast, wireless devices do not have the possibility to detect a collision while they are transmitting.
In fact, wireless stations can only learn about the success (or failure) of a transmission by means of feedback (or lack thereof) from the receiver.
For this reason, the length of a collision slot is approximately equal to the length of the longest transmission involved in the collision.

To reduce the likelihood of collisions, in CSMA/CA, the channel is divided into slots and transmissions are synchronized to slot borders and preceded by a random backoff.
In particular, stations performing backoff set a backoff counter to a randomly chosen value and decrement it by one at every slot.
The transmission occurs when the backoff counter reaches zero.

\subsection{The construction of a collision-free schedule for two contending stations}
CSMA/ECA is simply a subtle variant of the protocol described above.
The only difference between CSMA/CA and CSMA/ECA is that the latter uses a deterministic backoff after successful transmissions.
This deterministic backoff is constant and is the same for all the stations.
As a result, two stations that successfully transmit in two different slots will not collide with each other in their next transmission attempt.

Imagine that two stations STA 1 and STA 2 successfully transmit in two different slots ($X$ and $Y$),  and then they both backoff for the same number of slots $V$.
Their next transmission attempt occurs at slot $X+V$ and $Y+V$, which are different since $X$ and $Y$ are different.

The behavior of CSMA/CA and CSMA/ECA for a network of two nodes is depicted in Fig.~\ref{fig:ca_vs_eca}.
Fig.~\ref{fig:csma_ca} represents two stations competing for the channel using CSMA/CA.
The channel time is slotted and some slots are empty while others are busy with successes or collisions.
If realistic channels are considered, it is also possible that a busy slot contains a transmission that cannot be decoded due to unfavorable channel conditions.
Nevertheless, in this tutorial introduction, we will consider only an ideal channel that does not introduce errors.

The figure is not to scale for the ease of representation.
In reality, the busy slots are much longer than the empty ones. 
The figure also shows the backoff value of each of the two competing stations in each slot, and the tiny arrows indicate whether the backoff is randomly or deterministically selected.

It can be observed that the backoff value is decremented by one in every slot and that a station transmits when its backoff counter reaches zero.
After a transmission, each CSMA/CA station randomly chooses a new backoff value.

In the present example we assume that, after completing a transmission, each station has another packet to transmit.
In the literature, this particular assumption is often referred to as saturation condition (e.g., \cite{he2009srb,barcelo2010fcc,fang2011dlm,barcelo2011tcf}).
We will keep the saturation assumption in the remainder of the paper, although in the next section we will mention references that address the non-saturation scenario.

The CSMA/CA stations in Fig.~\ref{fig:csma_ca} always use a random backoff, which means that they are always exposed to a collision probability greater than zero.
It is useful to compare the behavior of CSMA/CA in Fig.~\ref{fig:csma_ca} to the behavior of CSMA/ECA in Fig.~\ref{fig:csma_eca}.
The initial behavior is exactly the same for the two protocols: a collision occurs and a random backoff is selected.
However, after the first successful transmission of STA 1 we can observe that the CSMA/ECA station deterministically chooses its backoff value.
The same occurs after the first successful transmission of STA 2.
The fact that the stations have successfully transmitted in different slots and use the same deterministic backoff value guarantees that these two stations will not collide with each other in their next transmission attempt.
From this point on, the behavior of the system is collision-free, deterministic, cyclic and fair.
The cycle length is indicated in the figure, and it is easy to observe that the behavior of the system in the second cycle is exactly the same as in the first cycle.
There is no need for a global agreement about which is the first slot of a cycle.
For example, each station can consider its own transmitting slot as the first slot of the cycle.

\subsection{Generalization to a larger number of contenders}
\begin{figure*}[!t]
\centering
\subfigure[CSMA/CA]{\includegraphics[width=2.5in]{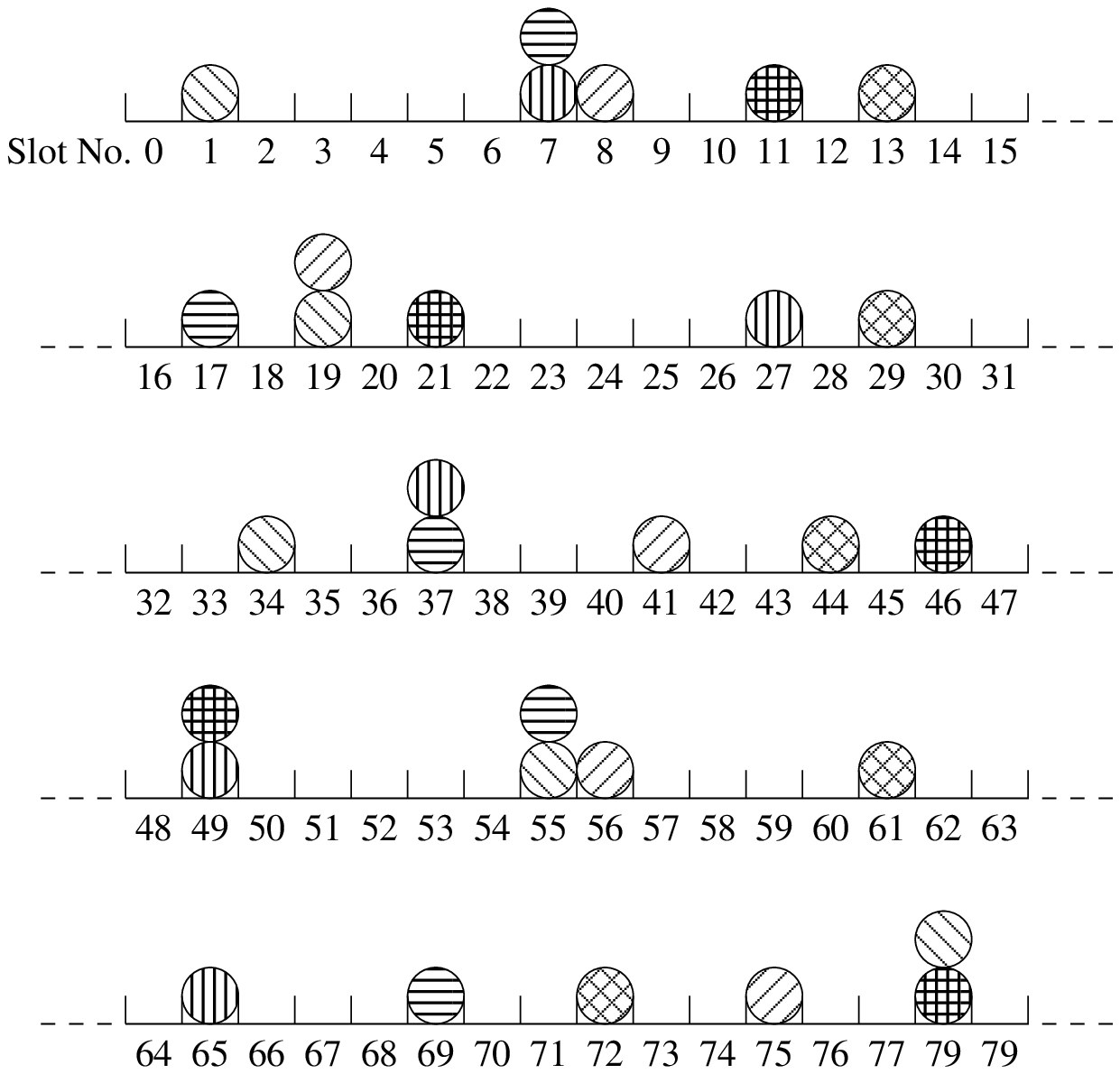}%
\label{fig:csma_ca_compact}}
\hspace{25mm}
\subfigure[CSMA/ECA]{\includegraphics[width=2.5in]{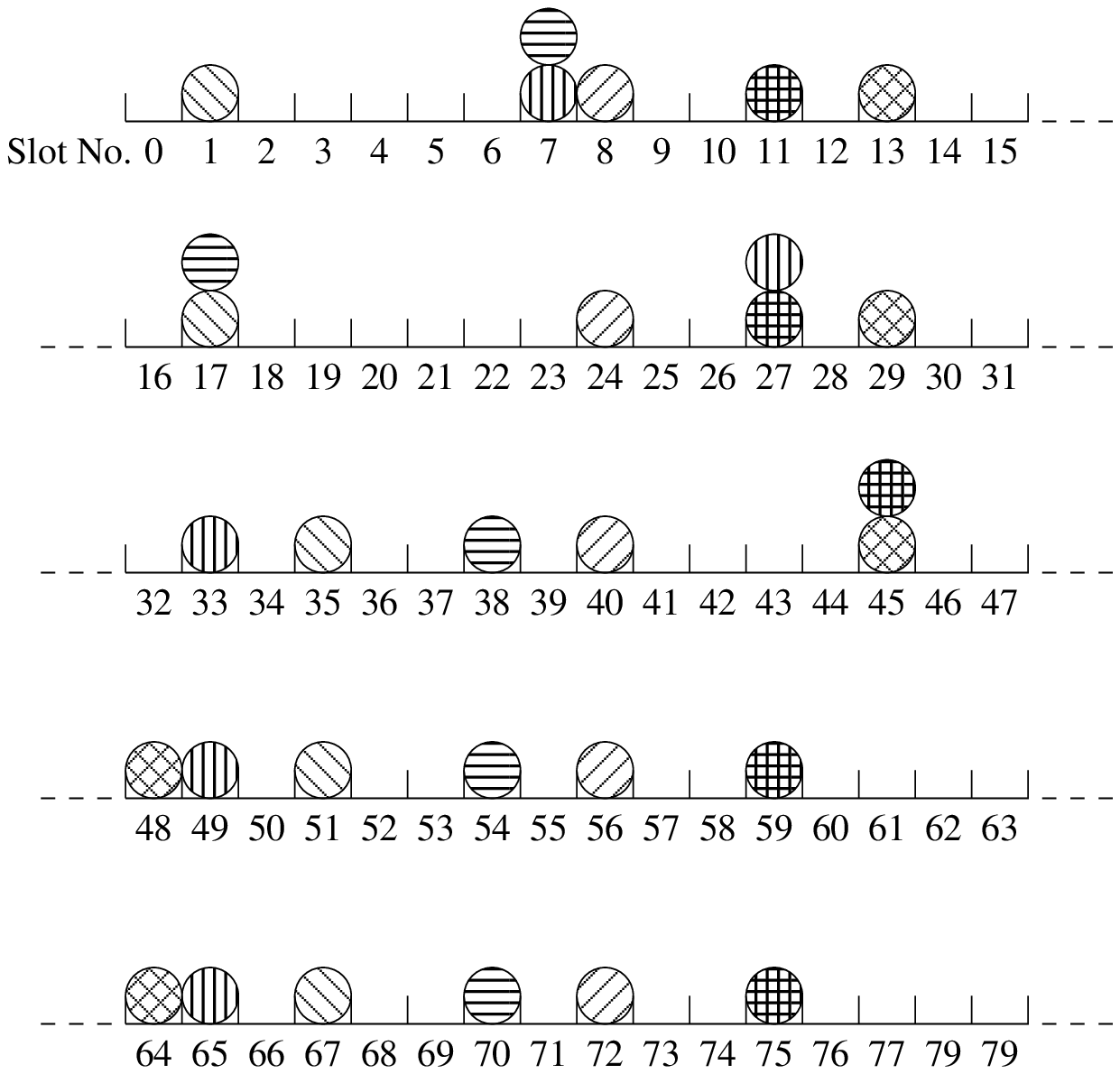}%
\label{fig:csma_eca_compact}}
\caption{A compact representation of contention in which six wireless stations compete for channel access. The disks represent the transmissions of the stations and the  patterns are used to identify the station that transmitted. The construction of the collision-free schedule in CSMA/ECA finishes when all the stations successfully transmit in the same cycle.}
\label{fig:ca_vs_eca_compact}
\end{figure*}

The general rule is that collision-free operation is reached after all the contending stations successfully transmit within a single cycle duration.
To better understand the construction of the collision-free schedule, it is useful to look at an example with more than two contending stations.
In order to depict the contention for the channel when the number of contenders is high, we will need a more compact representation such as the one used in Fig.~\ref{fig:ca_vs_eca_compact}.
For convenience, we draw all the slots with equal length.
Each slot is numbered and the transmissions of the stations are represented as disks in the slots.
There are six different stations competing for the channel and the hatching pattern of each disk identifies the transmitting station.

As in the previous example in Fig.~\ref{fig:csma_eca}, in the CSMA/ECA example in Fig.~\ref{fig:csma_eca_compact} the stations use a deterministic backoff after successful transmissions.
For convenience, the slots have been arranged in such a way that a deterministic backoff is represented by a new transmission in the same column of the following row.
As an example, the CSMA/ECA station that successfully transmits in slot 1 transmits again in slot 17, in the same column.
If we focus on the two CSMA/ECA stations that collide in slot 7, we realize that they use a random backoff which means that the new transmissions will probably end up in a different column.
In this particular example, the colliding stations in slot 7 retransmit in slot 17 and 27.
In CSMA/ECA, when all the stations successfully transmit in the same cycle, they all stick to the same column.
At this point, the collision-free schedule has already been constructed as we can observe in the last two rows of Fig.~\ref{fig:csma_eca_compact}.

The construction of the collision-free schedule results in significant performance gains in terms of throughput, as we will see in the next section.
Because the deterministic stations may only collide with random stations and not with one another, CSMA/ECA delivers a performance advantage even before the collision-free schedule is completely constructed.
This means that CSMA/ECA also outperforms CSMA/CA in highly dynamic scenarios in which the stations join and leave the contention.
In the extreme case in which the stations join the contention to transmit a single packet and then they leave, the performance of CSMA/ECA falls back to that of CSMA/CA.

A key aspect of the proposed protocol is that of the schedule length, which is equivalent to the deterministic backoff used after successful transmissions.
If the schedule length is excessively large compared to the number of contenders, the large number of empty slots will slightly penalize the performance.
On the other hand, if the schedule is too short, it will not be possible to accommodate the collision-free operation of all the participants.
As pointed out in \cite{fang2011dlm}, having a schedule that is larger than the number of contenders is better than having one that is shorter.
The reason is that empty slots are much shorter than collision slots and therefore, idle waiting is far less costly than collisions.
We discuss in the next section the possibility to adapt the schedule length in a distributed way.

Even though there are clear similarities between CSMA/ECA and Reservation Aloha, there are also two remarkable differences.
The first one is that in Reservation Aloha the slot size is fixed, while in CSMA/ECA the slot size is variable.
The second difference is that in Reservation Aloha there is slot reservation while in CSMA/ECA there is not.
A station that successfully transmits in CSMA/ECA can suffer a collision in its next transmission attempt, because there is no reservation in place.
Since there is no reservation, a station behaving randomly may choose the same slot as a station that is behaving deterministically.

The lack of reservations in CSMA/ECA makes the protocol very similar to CSMA/CA and allows for the peaceful coexistence of both protocols in the same network.
The similarity of CSMA/CA and CSMA/ECA is also an advantage as it eases the adaptation of current designs to the new protocol.
It is remarkable that the performance advantage of CSMA/ECA does not come at the price of additional signaling or extra overheads.

This section has covered the basic idea that enables the construction of a collision-free schedule in a highly idealized and simplified scenario. If CSMA/ECA is to be considered as a replacement of CSMA/CA, wider and deeper analysis is needed. The following section offers an overview of some contributions in this particular research area.

\section{Mathematical framework, performance evaluation, and refinements}
\label{sec:survey}
In this section we will summarize a small subset of representative contributions to offer an overview of some of the problems and possible enhancements of the basic idea described in the previous section. 

\subsection{Underlying mathematical framework}
Even though CSMA/ECA was initially suggested to prevent collisions in WLANs, the underlying mathematical framework is applicable to various resource allocation problems in the field of wireless networking, such as cognitive radio \cite{khan2013aso}, channel selection and network coding \cite{duffy2011dcs}.
The construction of a collision-free schedule in CSMA/ECA is in fact just an instance of a Constraint Satisfaction Problem (CSP) that the participating entities need to solve without explicit communication.
It is proven in \cite{duffy2011dcs} that the stochastic decentralized CSP solver (which is a generalization of the protocol that we have introduced in the previous section) guarantees that a solution will be found in finite time, if a solution exists.
Furthermore, its performance is competitive with some of the well-known centralized CSP solvers.

\subsection{Distributed adjustment of the cycle length}

Some improvements on the basic idea described in Sec.~\ref{sec:eca} are presented in \cite{fang2011dlm}. 
Namely, it suggests a distributed approach for adjusting the schedule length to accommodate a large number of contenders.
Furthermore, it introduces the concept of stickiness, whereby the stations stick to a deterministic backoff even after a transmission failure, for increased schedule robustness. 

Ideally, the deterministic backoff (which is equivalent to the number of columns in our representation) would be adjusted as a function of the number of contenders.
However, reaching this goal in a distributed fashion without requiring any kind of message exchange and preserving the system's fairness is quite a challenge.
The solution proposed in \cite{fang2011dlm} is elegant and effective:
A station that perceives a high collision probability doubles the deterministic backoff that it uses after successful transmissions.

\begin{figure*}[!t]
\centering
\includegraphics[width=6.0in]{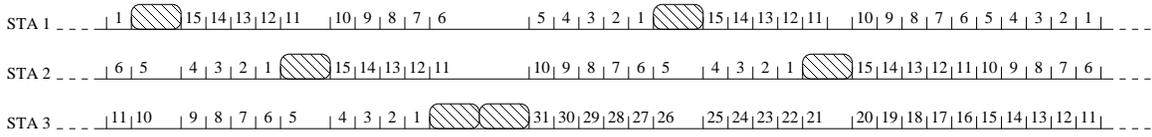}
\caption{Schedule length distributed adaption example. All the stations transmit the same number of packets in each cycle, despite using different schedule lengths.}
\label{fig:csma_eca_different_backoff}
\end{figure*}

The beautiful aspect of this approach is that the station that doubles its deterministic backoff also doubles the number of packets that are transmitted every time that it accesses the medium.
Using this trick, the number of available slots increases without any reduction in throughput.
In the long term, all the stations transmit the same number of packets, independently of their schedule length.
This property makes it possible for different stations to independently adjust their deterministic backoff value while preserving fairness.

As an example, consider the 3-node network shown in Fig.~\ref{fig:csma_eca_different_backoff}, where all three stations have reached a collision-free schedule.
Notice that the schedule length of STA 3 is twice as long as that of the other two stations.
Nevertheless, in terms of fairness, all the stations fairly share the channel.
The stations with short schedules transmit a single packet and the station with the long schedule transmits two packets when it is its turn.
Transmitting more than one packet when accessing the channel is possible and the latest revision of the IEEE 802.11 standard includes the necessary mechanisms for transmitting two or more packets back-to-back (packet aggregation).

\begin{figure}[!t]
\centering
\includegraphics[width=2.5in]{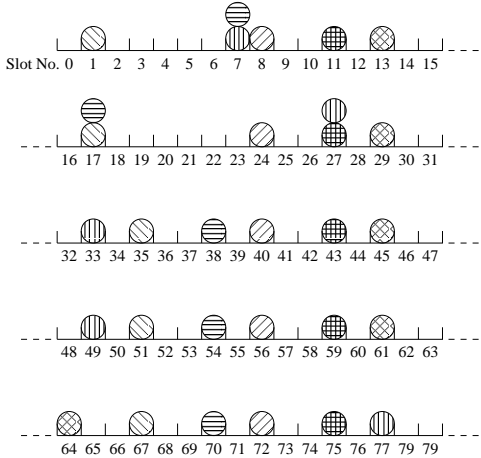}
\caption{The use of a deterministic backoff for two consecutive times after each successful transmission (CSMA/E2CA) speeds up the construction of the collision-free schedule.}
\label{fig:csma_e2ca}
\end{figure}

\subsection{Stickiness for faster convergence and increased robustness}

Under ideal conditions, a station using a deterministic backoff may collide only with a station using a random backoff.
If a deterministic station collides with a random station, only one of them needs to choose a random backoff to prevent a new collision of the two stations in their next transmission attempt.
In fact, switching the deterministic station to random behavior only increases the chance of collision with other deterministic stations.
Consequently, the protocol might be improved if deterministic stations kept using a deterministic backoff even after collisions.
The property of using a deterministic backoff after suffering collisions is called ``stickyness'' \cite{fang2011dlm}.

In ideal conditions, this solution has several advantages.
Firstly, it converges faster to collision-free operation, as a station that successfully transmits once never switches back to the random behavior.
Secondly, once a collision-free schedule is built, it is not possible for a channel error to move the system back to the random behavior.
And thirdly, after a collision-free schedule has been built, it is not possible for a new entrant to destroy the schedule.
The new entrant will simply use a random behavior (possibly suffering collisions) until it successfully transmits.

The problem is that real clocks may suffer drifts, that can result in slot misalignment \cite{gong2012asd}.
Two stations using a deterministic backoff may collide if their slot boundaries are not aligned.
This is a very undesirable situation if both stations ``stick'' to the use of deterministic backoff after colliding, as they will likely collide again in the next transmission attempt.

To benefit from the advantages of stickiness while preventing the aforementioned potential pitfall, \cite{fang2011dlm} proposes probabilistic stickiness and \cite{barcelo2011tcf} proposes finite stickiness, in which deterministic stations switch back to random behavior after a given number of consecutive collisions.
CSMA/E2CA in \cite{barcelo2011tcf} moves back to random behavior after two consecutive collisions.

Fig.~\ref{fig:csma_e2ca} illustrates the operation of the protocol when deterministic stations ``stick'' to a deterministic backoff after suffering a collision.
It can be observed that in this example, the convergence to collision-free operation is faster than in Fig.~\ref{fig:csma_eca_compact}.

\subsection{Performance of CSMA/ECA and CSMA/E2CA}

An analytical model of the expected number of slots required to reach collision-free operation is introduced in \cite{he2009srb}.
The paper also presents a comprehensive simulation study which includes realistic ingredients, such as traffic differentiation, carrier-sense errors, and channel errors.
Different performance metrics such as throughput, delay, and collision probability are evaluated, and both saturated and non-saturated traffic is considered.
The authors conclude that a protocol that uses a deterministic backoff after successful transmissions always outperforms the purely random protocol.
Interestingly, the authors report that the implementation of the proposed, protocol in the well-known simulator NS-2 required the change of only three lines of code.
This gives an idea of how similar the proposed protocol is to the legacy one,  and how easy it would be to include the proposed protocol in new devices.

Many performance aspects are covered in \cite{fang2011dlm}, which offers a comparison among different protocols that converge to collision-free operation, and studies the speed of convergence and performance in unsaturated scenarios and in the presence of errors and legacy stations.

Fairness of CSMA/ECA with regard to legacy stations is addressed in \cite{barcelo2010fcc}.
The results show that both protocols are interoperable and can fairly coexist in the same network.
CSMA/ECA stations will experience a slightly better performance than CSMA/CA stations, and the participation of legacy stations prevents the construction of a collision-free schedule.
Nevertheless, it is remarkable that the mix of new and legacy stations attains a better performance than a network in which all the stations follow the legacy protocol.

Backward compatibility is of paramount importance for any improvement to be adopted in WLANs, since there is a large base of deployed hardware that will not be thrown away overnight.
The possibility of CSMA/ECA to peacefully coexist with the previous protocol ensures a smooth transition from one protocol to the other, with a coexistence period in which both protocols will interoperate.

\begin{figure}[!t]
\centering
\includegraphics[width=\linewidth]{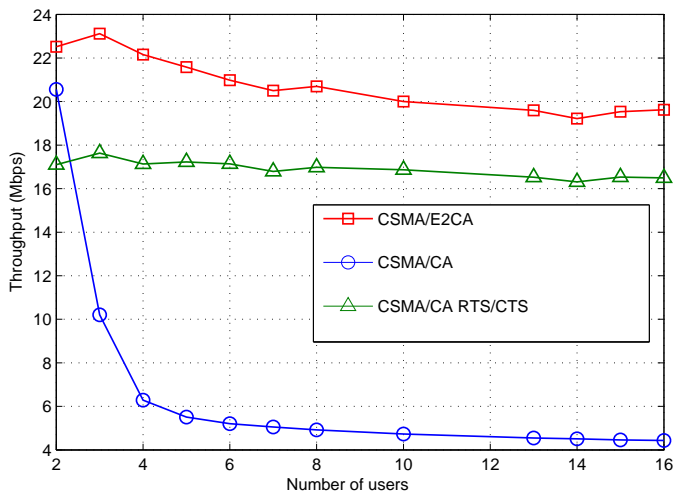}
\caption{Performance curves of CSMA/CA and CSMA/E2CA for an increasing number of contenders.}
\label{fig:performance}
\end{figure}

A detailed performance evaluation of CSMA/ECA is offered in \cite{martorell2012pec}.
This paper presents an analytical model and simulations that use realistic channel realizations and Automated Rate Fallback (ARF).
For comparison, results are also presented for CSMA/CA with and without the Request-To-Send/Clear-To-Send (RTS/CTS) four-way handshake.

ARF is a mechanism used to adapt the transmission rate to the channel conditions and simply works by reducing the transmission rate after unsuccessful transmissions.
This approach does not work well with CSMA/CA when collisions occur, as ARF misinterprets all failures as channel errors and reduces the transmission rate, which further worsens the performance when failures are due to collisions.
This problem can be alleviated by using RTS/CTS that can differentiate between collisions and channel errors.
However RTS/CTS penalizes the performance due to the additional overheads.
CSMA/ECA solves the problem by preventing collisions, without adding any additional overhead.
The curves in Fig.~\ref{fig:performance} (reproduced from \cite{martorell2012pec}) show that CSMA/ECA outperforms CSMA/CA and also CSMA/CA with RTS/CTS.

\section{Conclusion} \label{sec:conclusion}
In this paper we have summarized a family of MAC protocols that offer significant performance improvement over CSMA/CA. In its most basic form, CSMA/ECA achieves this performance boost by simply using a deterministic backoff after each successful transmission and reverting back to the CSMA/CA random behavior when a collision is detected. Under certain conditions, this leads to the construction of a collision-free deterministic schedule in a completely distributed fashion. We have then discussed possible variations to CSMA/ECA that can further improve the performance under more realistic conditions. CSMA/ECA and its variants represent a simple evolution of the currently prevalent protocol CSMA/CA, thus offering backward compatibility and fair coexistence with already deployed hardware. 

Similar techniques can be used to address other problems in wireless networking, such as channel assignment, spreading
code assignment, and channel sensing-order assignment in cognitive radio.

\bibliographystyle{IEEEtran}
\bibliography{IEEEabrv,my_bib}

\begin{thebibliography}{10}
\providecommand{\url}[1]{#1}
\csname url@samestyle\endcsname
\providecommand{\newblock}{\relax}
\providecommand{\bibinfo}[2]{#2}
\providecommand{\BIBentrySTDinterwordspacing}{\spaceskip=0pt\relax}
\providecommand{\BIBentryALTinterwordstretchfactor}{4}
\providecommand{\BIBentryALTinterwordspacing}{\spaceskip=\fontdimen2\font plus
\BIBentryALTinterwordstretchfactor\fontdimen3\font minus
  \fontdimen4\font\relax}
\providecommand{\BIBforeignlanguage}[2]{{%
\expandafter\ifx\csname l@#1\endcsname\relax
\typeout{** WARNING: IEEEtran.bst: No hyphenation pattern has been}%
\typeout{** loaded for the language `#1'. Using the pattern for}%
\typeout{** the default language instead.}%
\else
\language=\csname l@#1\endcsname
\fi
#2}}
\providecommand{\BIBdecl}{\relax}
\BIBdecl

\bibitem{abramson2009asw}
N.~Abramson, ``{The Alohanet - Surfing for Wireless Data [History of
  Communications]},'' \emph{{IEEE} Commun. Mag.}, vol.~47, no.~12, pp. 21--25,
  2009.

\bibitem{crowther1973sbc}
W.~Crowther, R.~Rettberg, D.~Walden, S.~Ornstein, and F.~Heart, ``{A System for
  Broadcast Communication: Reservation-ALOHA},'' in \emph{Hawaii Int. Conf.
  Syst. Sci}, 1973, pp. 596--603.

\bibitem{tasaka1983spr}
S.~Tasaka, ``{Stability and performance of the R-ALOHA packet broadcast
  system},'' \emph{Computers, IEEE Transactions on}, vol. 100, no.~8, pp.
  717--726, 1983.

\bibitem{barcelo2008lba}
J.~Barcelo, B.~Bellalta, C.~Cano, and M.~Oliver, ``{Learning-BEB: Avoiding
  Collisions in WLAN},'' in \emph{Eunice}, 2008.

\bibitem{he2009srb}
Y.~He, R.~Yuan, J.~Sun, and W.~Gong, ``{Semi-Random Backoff: Towards Resource
  Reservation for Channel Access in Wireless LANs},'' in \emph{IEEE ICNP},
  2009, pp. 21--30.

\bibitem{barcelo2010fcc}
J.~Barcelo, A.~Lopez-Toledo, C.~Cano, and M.~Oliver, ``{Fairness and
  Convergence of CSMA with Enhanced Collision Avoidance},'' in \emph{IEEE ICC},
  2010.

\bibitem{fang2011dlm}
M.~Fang, D.~Malone, K.~Duffy, and D.~Leith, ``{Decentralised Learning MACs for
  Collision-free Access in WLANs},'' \emph{Wireless Networks}, to appear.
  Available in SpringerLink.

\bibitem{barcelo2011tcf}
J.~Barcelo, B.~Bellalta, C.~Cano, A.~Sfairopoulou, and M.~Oliver, ``{Towards a
  Collision-Free WLAN: Dynamic Parameter Adjustment in CSMA/E2CA},'' in
  \emph{EURASIP Journal on Wireless Communications and Networking}, 2011.

\bibitem{martorell2012pec}
G.~Martorell, F.~Riera, G.~Femenias, J.~Barcelo, and B.~Bellalta, ``{On the
  performance evaluation of CSMA/E2CA protocol with open loop ARF-based
  adaptive modulation and coding},'' in \emph{European Wireless}, 2012.

\bibitem{khan2013aso}
Z.~Khan, J.~Lehtomaki, L.~DaSilva, and M.~Latva-aho, ``{Autonomous Sensing
  Order Selection Strategies Exploiting Channel Access Information},''
  \emph{Mobile Computing, IEEE Transactions on}, to appear. Available in IEEE
  Xplore.

\bibitem{duffy2011dcs}
K.~R. Duffy, C.~Bordenave, and D.~J. Leith, ``{Decentralized Constraint
  Satisfaction},'' \emph{Networking, IEEE transactions on}, to appear.
  Available in IEEE Xplore.

\bibitem{gong2012asd}
W.~Gong and D.~Malone, ``{Addressing Slot Drift in Decentralized Collision Free
  Access Schemes for WLANs},'' \emph{Multiple Access Communications}, pp.
  146--157, 2012.

\end{thebibliography}
\end{document}